\documentclass[useAMS,usenatbib]{mn2e}

\usepackage{graphicx} 
\usepackage{txfonts}
\usepackage{lscape}
\usepackage{mathrsfs}
\usepackage{nicefrac}

\title[Dark matter-rich early-type galaxies in the CASSOWARY 5 strong lensing system]{Dark matter-rich early-type galaxies in the CASSOWARY 5 strong lensing system}
\author[C. Grillo and L. Christensen]{C. Grillo$^{1}$\thanks{E-mail: grillo.claudio@googlemail.com} and L. Christensen$^{1}$\\
$^{1}$Excellence Cluster Universe, Technische Universit\"at M\"unchen, Boltzmannstr. 2, D-85748, Garching bei M\"unchen, Germany}

\begin{document}

\date{Accepted. Received; in original form}

\pagerange{\pageref{firstpage}--\pageref{lastpage}} \pubyear{2011}

\maketitle

\label{firstpage}

\begin{abstract}
We study the strong gravitational lensing system number 5 identified by the CAmbridge Sloan Survey Of Wide ARcs in the skY. In this system, a source at redshift 1.069 is lensed into four detected images by two early-type galaxies at redshift 0.388. The average projected angular distance of the multiple images from the primary lens is 12.6 kpc, corresponding to approximately 1.3 times the value of the galaxy effective radius. The observed positions of the multiple images are well reproduced by a model in which the total mass distribution of the deflector is described in terms of two singular isothermal sphere profiles and a small external shear component. The values of the effective velocity dispersions of the two lens galaxies are $328_{-8}^{+7}$ and $350_{-18}^{+17}$ km s$^{-1}$. The best-fit lensing model predicts magnification values larger than 2 for each multiple image and a total magnification factor of 17. By modelling the lens galaxy spectral energy distributions, we measure lens luminous masses of $(3.09\pm0.30)\times 10^{11}$ and $(5.87\pm0.58)\times 10^{11}$ $M_{\odot}$ and stellar mass-to-light ratios of $2.5 \pm 0.3$ and $2.8 \pm 0.3$ $M_{\odot}L_{\odot,i}^{-1}$ (in the observed $i$-band). These values are used to disentangle the luminous and dark matter components in the vicinity of the multiple images. We estimate that the dark over total mass ratio projected within a cylinder centred on the primary lens and with a radius of 12.6 kpc is $0.8\pm 0.1$. Inside the effective radii of the two galaxies, we measure projected total mass-to-light ratios of $12.6 \pm 1.4$ and $13.1 \pm 1.7$ $M_{\odot}L_{\odot,i}^{-1}$. We contrast these measurements with the typical values found at similar distances (in units of the effective radius) in isolated lens galaxies and show that the amount of dark matter present in these lens galaxies is almost a factor four larger than in field lens galaxies with comparable luminous masses. Data and models are therefore consistent with interpreting the lens of this system as a galaxy group. We infer that the overdense environment and dark matter concentration in these galaxies must have affected the assembly of the lens luminous mass components, resulting in the large values of the galaxy effective radii. We conclude that further multi-diagnostics analyses on the internal properties of galaxy groups have the potential of providing us a unique insight into the complex baryonic and dark-matter physics interplay that rules the formation of cosmological structures.
\end{abstract}


\section{Introduction}

\begin{table*}
\centering
\caption{The lens galaxies.}
\label{tab1}
\begin{tabular}{cccccccccc} 
\hline\hline \noalign{\smallskip}
& R.A. & Dec. & $z_{l}$ & $\theta_{e,i}$ & $u\,^{b}$ & $g\,^{b}$ & $r\,^{b}$ & $i\,^{b}$ & $z\,^{b}$ \\
& (J2000) & (J2000) & & (\arcsec) & (mag) & (mag) & (mag) & (mag) & (mag) \\
\noalign{\smallskip} \hline \noalign{\smallskip}
L1 & 12:44:51.36 & 01:06:42.9 & $0.3877\,^{a}$ & 1.87$\pm$0.18 & 22.26$\pm$0.50 & 20.55$\pm$0.04 & 19.08$\pm$0.02 & 18.42$\pm$0.02 & 18.02$\pm$0.04 \\
L2 & 12:44:51.01 & 01:06:43.9 & 0.3877 & 2.88$\pm$0.23 & 22.25$\pm$0.79 & 20.37$\pm$0.06 & 18.40$\pm$0.02 & 17.85$\pm$0.01 & 17.67$\pm$0.04 \\
\noalign{\smallskip} \hline
\end{tabular}
\begin{list}{}{}
\item[$^a$]Assumed to be the same as for L2.
\item[$^b$]AB magnitudes.
\end{list}
\end{table*}

Galaxy groups are interesting astrophysical systems that bridge the gap between isolated, single galaxies and massive galaxy clusters. A large fraction of all galaxies reside in this special transition regime (e.g., \citealt{eke04}), where the effects of baryonic physics can be investigated. In fact, while in the inner region of galaxies the total mass distribution is dominated by baryons (e.g., \citealt{sag92}; \citealt{ger01}; \citealt{tho07}; \citealt{gri09}) and in clusters of galaxies dark matter is the main mass component outside their central cores (e.g., \citealt{jee05}; \citealt{biv06}; \citealt{san08}; \citealt{new09}), the interplay between baryonic and dark matter is expected to play a key role in shaping galaxy groups, that are composed by a few gravitationally bound galaxies. Despite the recognised cosmological relevance of studies in groups of galaxies, the physical properties of these systems are still relatively unexplored.

Galaxy groups at low and intermediate redshifts have been investigated by means of X-ray and optical diagnostics (e.g., \citealt{hel00,hel03}; \citealt{osm04}; \citealt{ras07}; \citealt{gas07}; \citealt{fas08}; \citealt{yan08}; \citealt{sun09}) and only more recently some gravitational lensing analyses in galaxy groups have also become available (e.g., \citealt{aug07}; \citealt{kub09}; \citealt{lim09,lim10}; \citealt{tu09}; \citealt{ver10}; \citealt{tha10}). The modest number of known groups of galaxies showing strong lensing features and the smallness of the weak lensing signal (e.g., \citealt{hoe01}; \citealt{par05}) produced by these objects make detailed lens modelling at these mass scales particularly valuable. Strong lensing, in combination with multiband photometry or galaxy dynamics, can be exploited to disentangle the group luminous and dark components, in the same way as in single galaxies and clusters of galaxies (e.g., \citealt{gri08b,gri09,gri10c}; \citealt{bar09}; \citealt{san08}; \citealt{new09}). In suitable lensing systems, the extended luminosity distribution of the multiple images has also allowed to estimate the halo truncation radii of specific group members, showing the importance of tidal stripping processes in the structural evolution of galaxies located in overdense environments (e.g., \citealt{suy10}; \citealt{ric10}; \citealt{don10}). Moreover, the fairly high lensing magnification factors achieved in galaxy groups have offered the possibility of studying in detail the physical properties (like the star formation rate, the chemical abundance and composition) of distant lensed galaxies (e.g., \citealt{pet00,pet02,pet10}; \citealt{tep00}; \citealt{bak04}; \citealt{sia08}).

By means of different observational and selection criteria, galaxy groups acting as strong gravitational lenses on background sources have been identified in the Cosmic Lens All-Sky Survey (CLASS; \citealt{mye03}; \citealt{bro03}), Strong Lensing Legacy Survey (SL2S; \citealt{cab07}), and CAmbridge Sloan Survey Of Wide ARcs in the skY (CASSOWARY; \citealt{bel09}). For example, the CLASS survey has discovered through radio observations one system composed of two images, separated by 4.56\arcsec\, in projection, of a radio-loud quasar lensed by a massive early-type galaxy with a satellite galaxy (\citealt{mck05,mck10}). The other two surveys have concentrated on using optical multicolor images (from the Canada France Hawaii Telescope Legacy Survey, CFHTLS, and Sloan Digital Sky Survey, SDSS) to find potential lensed (blue) objects located around groups of massive (red) galaxies. Several of these photometrically selected systems have been spectroscopically followed up to validate the lensing hypothesis and to estimate the redshifts of the lenses and sources.

In this work, we concentrate on the system number five (CSWA 5) of the CASSOWARY catalogue\footnote{http://www.ast.cam.ac.uk/research/cassowary/}. As for all systems in the CASSOWARY survey, \emph{ugriz} imaging from the SDSS is available. From a colour-composite image (see \citealt{chr10}), it is possible to distinguish two similar red galaxies (L1 and L2, at a projected angular distance of 5.27\arcsec), one of which has a SDSS spectrum showing absorption features redshifted at $z_{l}=0.3877$, and four faint blue images (im1-im4) at projected angular distances of a few arcsececonds (the lens galaxies and the multiple images are labelled in the second panel of Fig. \ref{fig1}). Three of these images have been observed with the X-shooter spectrograph at the Very Large Telescope (VLT) and have been verified to be the lensed multiple images of a star-forming galaxy located at redshift $z_{s}=1.0686$. The excellent spectral resolution and wavelength coverage of the X-shooter instrument, combined with the lensing magnification effect, have allowed a refined study of the physical properties of this distant source (see \citealt{chr10}). The study presented here focuses on the characterization of the lens galaxies and completes the previous analysis of this lensing system. In the following, we will provide realistic lensing modelling and measure the luminous and dark matter components of the deflector.

This paper is organized as follows. In Sect. 2, we use photometric redshift estimators to prove that the data for the two red galaxies are consistent with L1 and L2 being at the same redshift. In Sect. 3, we develop strong lensing models for the CSWA 5 system, under the assumptions that the two galaxies are both lenses at the same redshift $z_{l}=0.3877$ and that the four images are all multiple images of the same blue source at $z_{s}=1.0686$. Then, in Sect. 4, we fit the SDSS multiband photometry of the two lenses with composite stellar population templates to derive the galaxy luminous masses and compare these values to the total mass estimates provided by the strong lensing analysis. This piece of information is employed to estimate the amount and distribution of dark matter in the galaxies. Finally, in Sect. 5, we summarise our work and draw conclusions. Throughout this work we assume the following values for the cosmological parameters: $H_{0}=70$ km s$^{-1}$ Mpc$^{-1}$, $\Omega_{m}=0.3$, and $\Omega_{\Lambda}=0.7$. In this model, 1\arcsec$\ $ corresponds to a linear size of 5.27 kpc at the lens plane. 

\begin{table}
\centering
\caption{The lensed images.}
\label{tab2}
\begin{tabular}{cccccc} 
\hline\hline \noalign{\smallskip}
& $x_{i}\,^{a}$ & $y_{i}\,^{a}$ & $z_{s}$ & $\delta_{x,y}$ & $d\,^{a}$ \\
& (\arcsec) & (\arcsec) & & (\arcsec) & (\arcsec) \\
\noalign{\smallskip} \hline \noalign{\smallskip}
im1 & 1.92 & $-2.90$ & $1.0686\,^{b}$ & 0.10 & 3.48 \\
im2 & 1.18 & 2.20 & $1.0686\,^{b}$ & 0.10 & 2.50 \\
im3 & 1.77 & 1.20 & $1.0686\,^{b}$ & 0.10 & 2.14 \\
im4 & $-1.33$ & 0.50 & $1.0686\,^{c}$ & 0.10 & 1.42 \\
\noalign{\smallskip} \hline
\end{tabular}
\begin{list}{}{}
\item[$^a$]With respect to L1.
\item[$^b$]Measured by \citet{chr10}.
\item[$^c$]Assumed to be the same as for im1-im3.
\end{list}
\end{table}

\section{The hypothesis of a lens galaxy group}

The SDSS provides one of the largest multi-color ($ugriz$) imaging catalogue available to date. By using the SDSS five-band photometry, that covers a wavelength range from 354 to 913 nm, an approximate redshift (called photometric redshift or photo-z) of each galaxy in the survey can be estimated. We refer in the following to the three different photo-z estimators included in the SDSS database and identify them as $z_{ph_{1}}$, $z_{ph_{2}}$, and $z_{ph_{3}}$, respectively.

The photometric redshift estimates are all determined from dereddened total magnitudes, that provide unbiased galaxy colours in absence of colour gradients. The $z_{ph_{1}}$ estimator assigns a redshift to a galaxy by fitting its observed broadband filter fluxes with redshift-evolved spectral energy distribution (SED) templates, that are calibrated with a training sample of objects with photometric data and spectroscopic redshifts (for further details see \citealt{csa03} and references therein). The other two estimators employ the artificial neural network technique to derive empirically a relation between photometric observables and redshifts in a training set of photometrically and spectroscopically observed galaxies. From this relation, the redshifts of the galaxies with only photometric data can then be obtained. As input parameters, $z_{ph_{2}}$ uses the four colors $u-g$, $g-r$, $r-i$, and $i-z$ and the three concentration indices in the $gri$ filters, $z_{ph_{3}}$ the five $ugriz$ magnitudes and concentration indices (for definitions and details see \citealt{oya08}).

Template-fitting and empirical methods have different advantages and disadvantages. The former can be applied to galaxies with a wide range of redshifts and intrinsic colours and provide, in addition to redshift estimates, rest-frame magnitudes as well, the latter are independent from systematical errors in the photometric calibrations of the data. By contrast, template-fitting techniques require a large set of galaxy SEDs and are affected by systematics due to calibration, empirical techniques need a training set of galaxies with photometric and spectroscopic properties that resemble those of the galaxies in the sample to be studied. For these reasons, we consider here estimators of different kind.

In Table \ref{tab4}, we show that each of the three estimators $z_{ph_{1}}$, $z_{ph_{2}}$, and $z_{ph_{3}}$ provides for the two lens galaxies, L1 and L2, photometric redshift values that are consistent within the errors. Given the robustness of the considered photo-z estimators, that have been calibrated on several thousands of galaxies in the SDSS, in the remainder of this paper we will approximate the deflector with a thin lens formed by two main mass components located at the same distance. To test the hypothesis that L1 and L2 reside in a collapsed galaxy group, we will proceed in the following sections to construct a lens model and disentangle the luminous and dark matter components. A possible evidence of the deflector being a galaxy group would be a dark over total mass fraction in these lens galaxies larger than in isolated lens galaxies. We will use the spectroscopic redshift value available for L2 ($z_{l}=0.3877$) to estimate the angular diameter and luminosity distances that are relevant to model this strong lensing system.

\begin{table}
\centering
\caption{The photometric redshift estimates of the lens galaxies.}
\label{tab4}
\begin{tabular}{cccc} 
\hline\hline \noalign{\smallskip}
 & $z_{ph_{1}}$ & $z_{ph_{2}}$ & $z_{ph_{3}}$ \\
\noalign{\smallskip} \hline \noalign{\smallskip}
L1 & 0.36$\pm$0.03$\,^{a}$ & 0.35$\pm$0.03$\,^{b}$ & 0.38$\pm$0.03$\,^{b}$ \\
L2 & 0.36$\pm$0.03$\,^{a}$ & 0.35$\pm$0.02$\,^{b}$ & 0.32$\pm$0.02$\,^{b}$ \\
\noalign{\smallskip} \hline
\end{tabular}
\begin{list}{}{}
\item[$^a$]Data Release 8.
\item[$^b$]Data Release 7. For these estimators, Data Release 8 values are not yet available. 
\end{list}
\end{table}

\section{The strong gravitational lensing modelling}

\begin{figure*}
  \centering
  \includegraphics[width=0.49\textwidth]{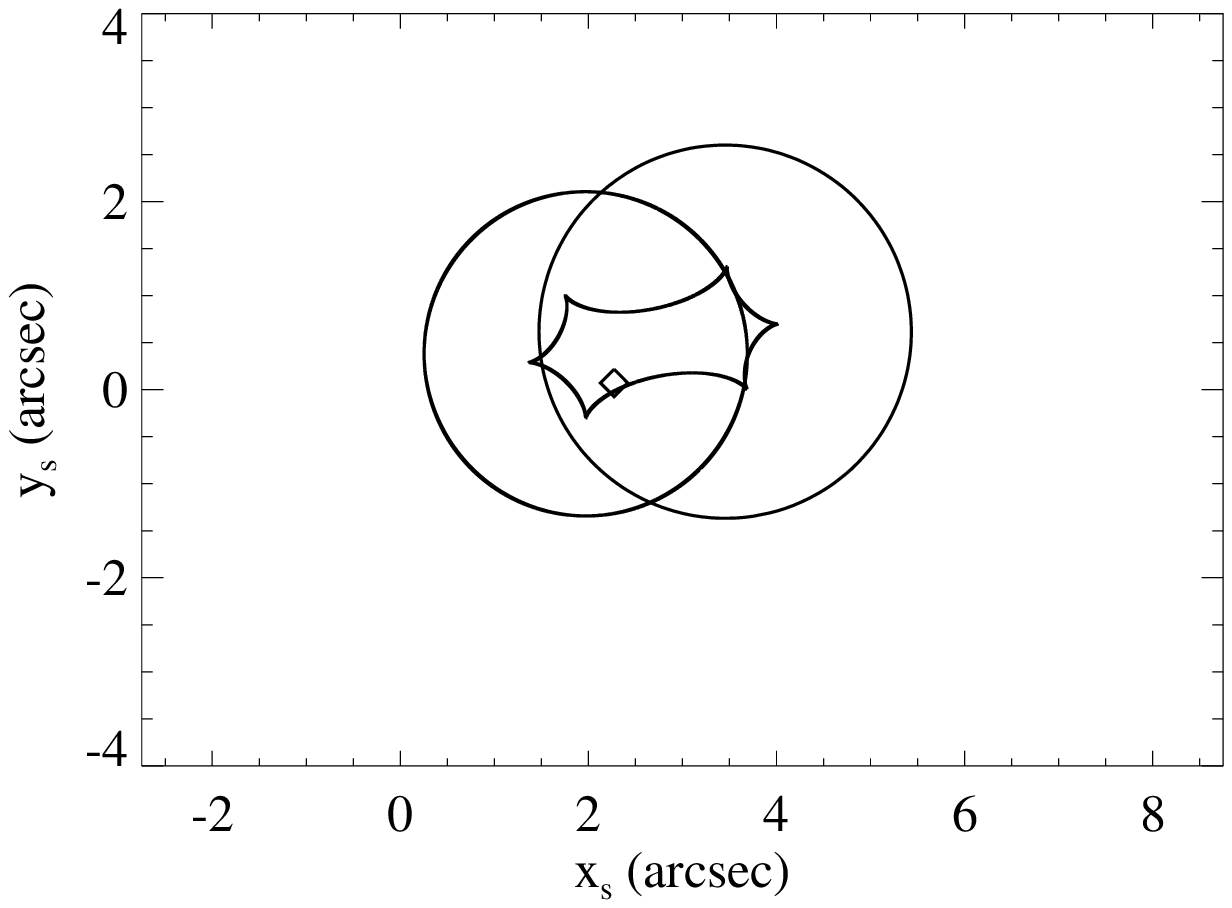}
  \includegraphics[width=0.49\textwidth]{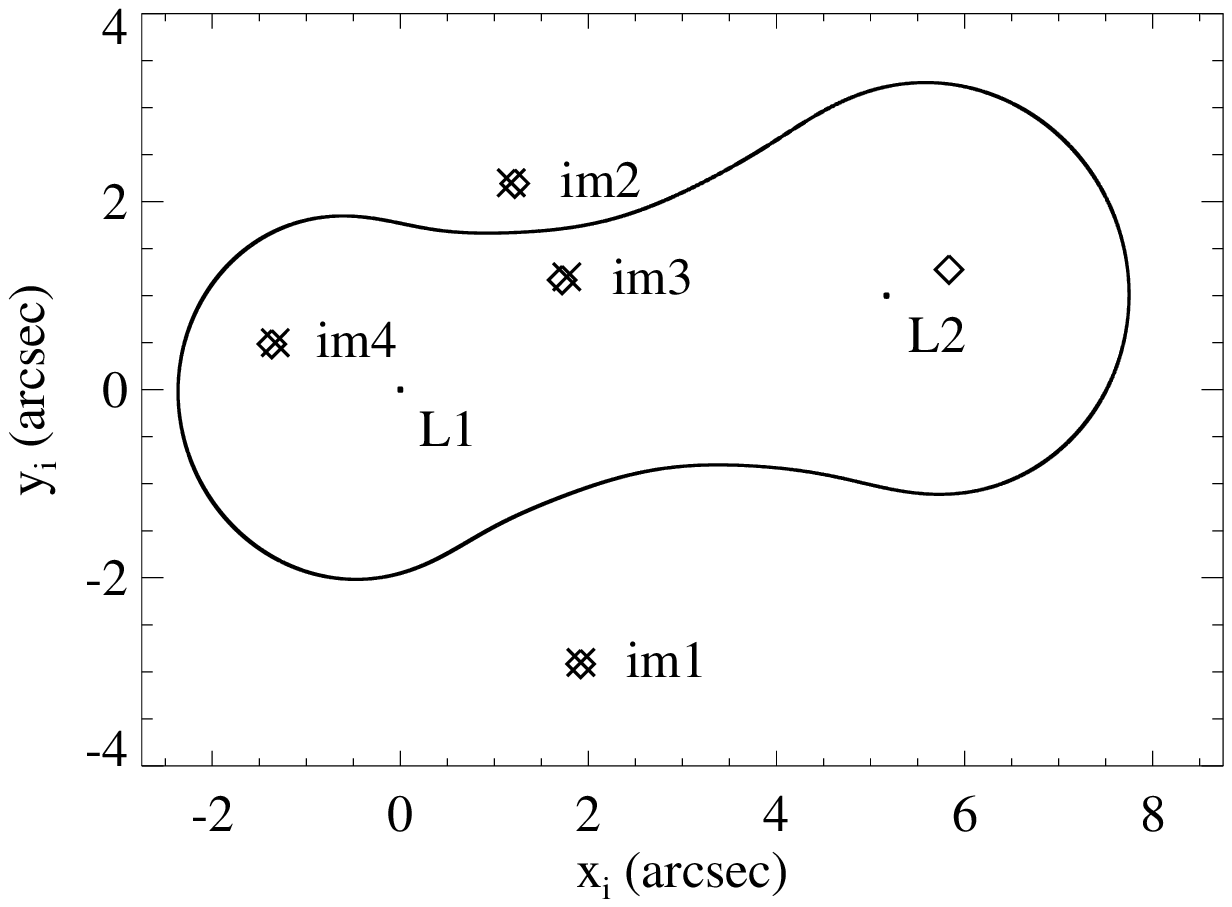}
  \includegraphics[width=0.49\textwidth]{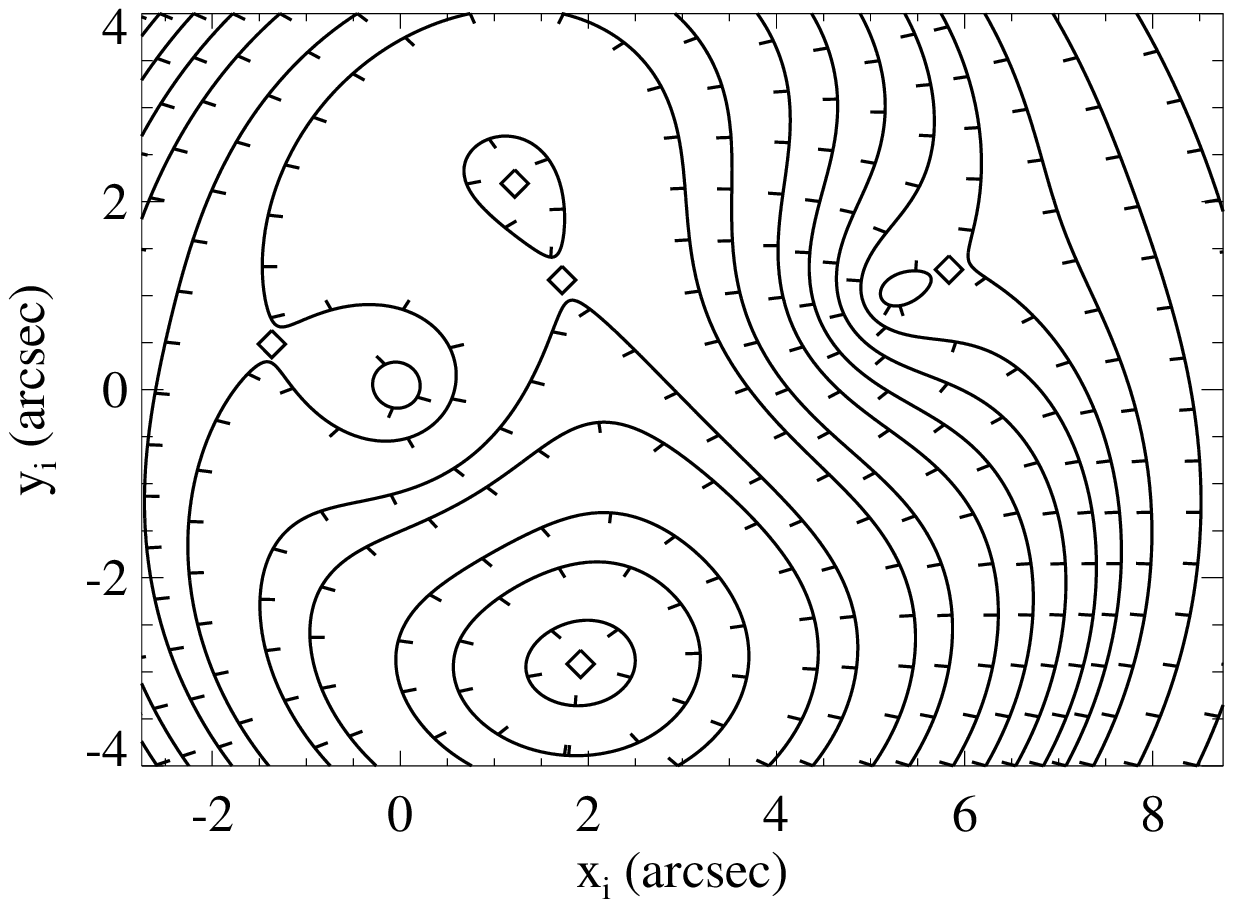}
  \includegraphics[width=0.49\textwidth]{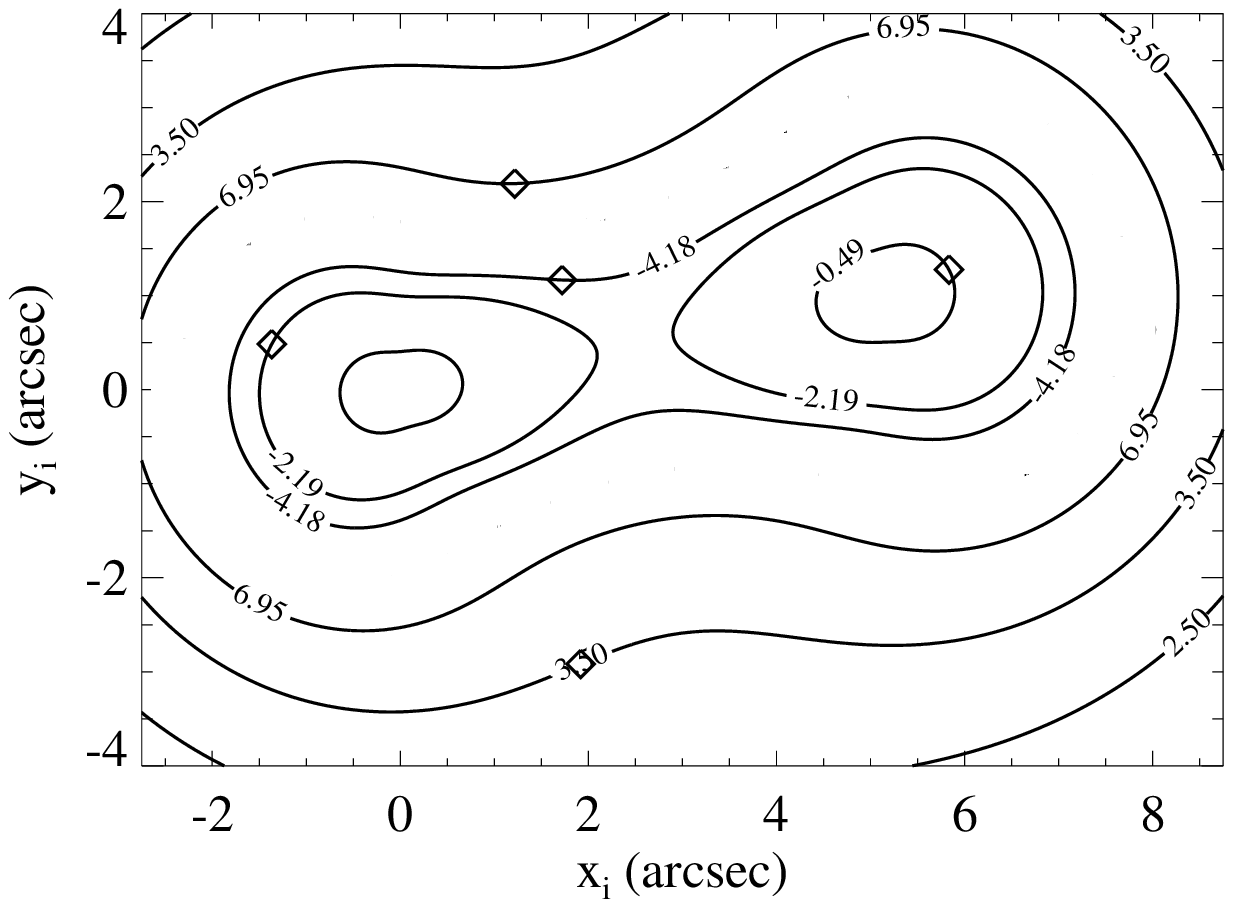}
  \caption{Best SIS+ES model. \emph{Top left:} The source plane with the caustics and the predicted position of the source (diamond symbol). \emph{Top right:} The image plane with the critical lines and the four observed (cross symbols) and five predicted (diamond symbols) positions of the images. \emph{Bottom left:} Contour levels of the Fermat potential showing two minima (im1 and im2) and three saddle points (im3, im4, and the additional fifth image). \emph{Bottom right:} Contour levels of the magnification factor with the values at the predicted positions of the multiple images. Positive and negative values represent, respectively, images with parity conserved or inverted with respect to the source.}
  \label{fig1}
\end{figure*}

In this section we use the public software Gravlens\footnote{http://redfive.rutgers.edu/$\sim$keeton/gravlens/} (\citealt{kee01a}) to perform a strong lensing analysis of the CSWA 5 system, focussing in particular on the determination of the total mass properties of the deflector.

We model the source and the multiple images as point-like objects and the projected total mass distribution of the main lenses (i.e., the galaxies L1 and L2 of Table \ref{tab1}) in terms of either axisymmetric de Vaucouleurs (deV) or singular isothermal sphere (SIS) profiles, without and with an external shear (ES) component (for more details on the models, see e.g. \citealt{kee01b}). These models are parametrised by the values of two length scales ($b_{L1/L2}$), related to the values of the Einstein angles of the lenses, and of the external shear intensity ($\gamma$) and position angle ($\theta_{\gamma}$). Given the quality of our deepest image (i.e., the X-shooter acquisition image shown in \citealt{chr10}, with an instrumental pixel size of 0.173\arcsec, we consider conservative positional errors, $\delta_{x,y}$, of 0.1\arcsec\, on the values of the centroids, $x_{i}$ and $y_{i}$, of the lensed images (see Table \ref{tab2}). We measure that the average value of the angular distances $d$ of the multiple images from the centre of the primary lens L1 is 2.38\arcsec\, (corresponding to 12.6 kpc). We fix the centres of mass of the lenses to their centres of light and, for the deV models, the effective angles $\theta_{e}$ to the values of Table \ref{tab1}, obtained by fitting de Vaucouleurs profiles to the SDSS \emph{i}-band luminosity distributions of the galaxies and taking appropriately into account the point spread function (PSF). We minimise a standard chi-square $\chi^{2}$ function between the observed and model-predicted positions of the multiple images over the model parameters and compare the best-fit (minimum) $\chi^{2}$ values to the number of degrees of freedom (d.o.f.). The latter is given by the difference between the numbers of the observables (eight coordinates identifying the positions of the four images) and that of the model parameters (two coordinates locating the source position plus the values of $b_{L1}$ and $b_{L2}$ and, when present, those of $\gamma$ and $\theta_{\gamma}$). The best-fit (minimum chi-square) parameters are listed in Table \ref{tab3}.

\begin{table}
\centering
\caption{The best-fit (minimum chi-square) models.}
\label{tab3}
\begin{tabular}{lcccccc} 
\hline\hline \noalign{\smallskip}
Model & $b_{L1}$ & $b_{L2}$ & $\gamma$ & $\theta_{\gamma}$ & $\chi^{2}$ & d.o.f. \\
& (\arcsec) & (\arcsec) & & (deg) & & \\
\noalign{\smallskip} \hline \noalign{\smallskip}
deV & 4.27 & 5.22 & & & 4.17 & 4 \\
SIS & 1.76 & 1.96 & & & 1.41 & 4 \\
deV$+$ES & 4.23 & 5.18 & 0.03 & 164.8 & 2.22 & 2 \\
SIS$+$ES & 1.75 & 2.01 & 0.01 & 139.1 & 0.69 & 2 \\
\noalign{\smallskip} \hline
\end{tabular}
\end{table}

As demonstrated by the small values of the best-fit $\chi^{2}$ compared to the number of degrees of freedom, all models reproduce well the observed image positions. This result suggests that with this level of uncertainty on the image positions there is no need to consider additional parameters to the relatively simple (axisymmetric with fixed centres) total mass models adopted for the deflector. The $\chi^{2}$ values show that SIS profiles are from 2 to 4 times more likely than deV profiles to describe the projected total mass distributions of the two lens galaxies. This provides some evidence that the projected total mass of the lenses is less concentrated than their light. The small intensity and the position angle of the external shear are consistent and rather well oriented with the possible effect and orientation of the few galaxies that have SDSS colours or photometric redshifts comparable to those of L1 and L2 and are angularly close to them.

In Fig. \ref{fig1}, we plot for the best-fit (and most likely) SIS$+$ES model the source plane with the caustics and the reconstructed position of the lensed source, the image plane with the critical curves and the observed and reconstructed positions of the multiple images, the predicted Fermat potential with its stationary points, and magnification map (for definitions, see e.g. \citealt{sch92}). We remark that in addition to the accurate reproduction of the four image positions, the image magnification factors are also in qualitative agreement with the observations. A more quantitative comparison between the observed and model-predicted image fluxes would require a deeper and better angularly resolved image of the lensing system.

We notice that the best-fit model predicts magnification factors larger than 2 for all observed multiple images, resulting in a total magnification factor of approximately 17, and a fifth demagnified image that is located angularly very close to the center of L2. We use GALFIT (\citealt{pen10}) on the X-shooter acquisition image to model the surface brightness distribution of L2 with a Sersic profile, keeping the centre fixed and fitting the values of Sersic index, effective angle, ellipticity, and its position angle. GALFIT includes a model of the PSF shape when fitting galaxy morphologies. However, in the 20 s integration time for our image there are only two stars in the field which are of the same magnitude or slightly fainter than L2, and therefore not optimal for modelling the PSF. After subtracting the model, the residuals within 1\arcsec\, $\ $from the centre of L2 are consistent with the noise in the frame. We conclude that deeper data are necessary to detect the possible presence of a fifth image of the source. We also scan the 3\arcsec -diameter fiber-spectrum of L2 from the SDSS but do not find any significantly high S/N emission line that can be associated to some light coming from the possible fifth image and entering in the fibre aperture. This is not worrying since this predicted additional image is expected to be more than four times fainter than the faintest observed image (im4). Moreover, we investigate the possibility of having a three instead of a five multiple-image system, as theoretically studied by \citet{shi08} in cases where two main mass concentrations are located on the same lens plane. We reconsider our lensing models with only the spectroscopically confirmed im1, im2, and im3. We find best-fit $\chi^{2}$ values on the order of $10^{3}$. This fact, combined with the good level with which the four image positions and magnifications can be reproduced, supports the hypothesis that this lensing system is more likely composed of a source that is lensed five times.

\begin{figure}
  \centering
  \includegraphics[width=0.49\textwidth]{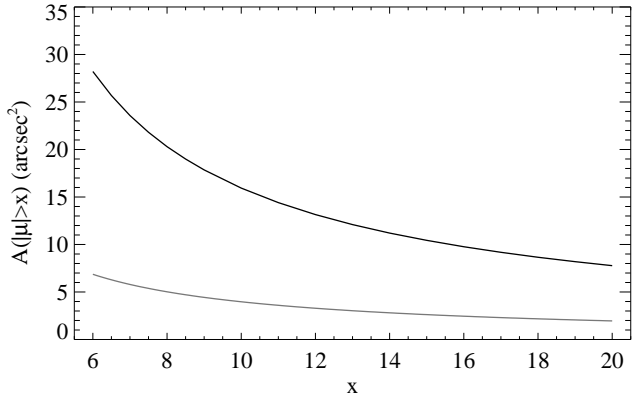}
  \caption{Area ($A$) on the image plane where the absolute value of the magnification factor ($\mu$) of the best-fit SIS+ES model is larger than a certain threshold ($x$) (in black). For comparison with single lens systems, the same function is plotted using the previous best-fit values, removing the lens galaxy L2, and considering only L1 (in grey).}
  \label{fig5}
\end{figure}

In Fig. \ref{fig5}, we show for the best-fit SIS+ES model the area on the image plane where the magnification factor $\mu$ has an absolute value that is larger than some fixed limits. We also compare the extensions of the highly magnified regions to those of lensing systems composed of isolated massive lens galaxies. By using the same best-fit model parameters and excluding L2, we find that L1 alone would produce the same magnification factors on approximately four times smaller areas. Looking at Fig. \ref{fig5}, we remark that deflectors composed by more than one mass concentration can act as efficient gravitational telescopes, providing large magnification factors on fairly extended areas of the sky. This fact is particularly relevant for astrophysical studies on high-redshift sources (e.g., \citealt{pet00,pet02,pet10}; \citealt{tep00}; \citealt{bak04}; \citealt{sia08}; \citealt{chr10}) that, without lensing, would otherwise not be observable with the present technology.

\begin{figure}
  \centering
  \includegraphics[width=0.35\textwidth]{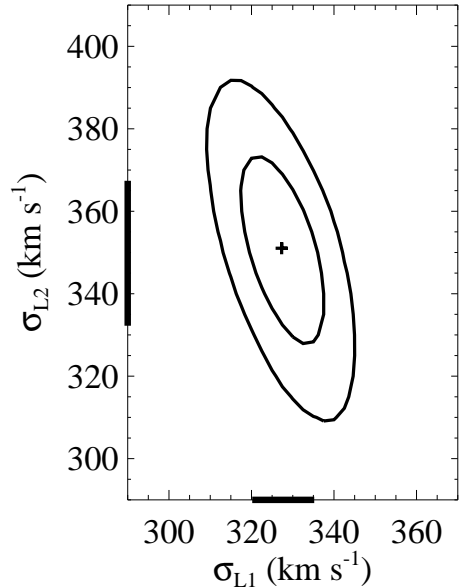}
  \caption{Estimates of the errors and correlation of the two lens effective velocity dispersions, $\sigma_{L1}$ and $\sigma_{L2}$, for the SIS+ES model. Results from a bootstrapping analysis performed on 2000 simulated data sets. The contour levels indicate the 68\% and 95\% confidence regions and the thick bars the 68\% confidence interval obtained by excluding the 320 smallest and largest best-fit values. A plus symbol locates the best-fit (minimum $\chi^{2}$) values of the two parameters corresponding to the values of $b_{L1}$ and $b_{L2}$ of Table \ref{tab3}.}
  \label{fig2}
\end{figure}

To quantify the statistical errors and degeneracy of the parameters $b_{L1}$ and $b_{L2}$, we perform a bootstrapping analysis by resampling the observed image positions according to their adopted positional errors and minimising the positional $\chi^{2}$ over the SIS+ES model parameters. We convert the best-fit values of $b_{L1}$ and $b_{L2}$ obtained from 2000 random data sets into effective velocity dispersion values $\sigma_{L1}$ and $\sigma_{L2}$ for the two lenses. We remind that for a SIS model the values of $b$ and $\sigma$ are related in the following way:
\begin{equation}
\label{eq1}
b=4 \pi \, \bigg(\frac{\sigma}{c} \bigg)^{2} \frac{D_{ls}}{D_{os}}\, ,
\end{equation}
where $c$ is the light speed and $D_{ls}$ and $D_{os}$ are, respectively, the angular diameter distances between the lens and the source and the observer and the source. We plot in Fig. \ref{fig2} the results of this analysis. 

We observe that the values of the two parameters $\sigma_{L1}$ and $\sigma_{L2}$ are clearly anticorrelated. This follows from the fact that the geometrical configuration of the multiple images provides a very precise estimate of the projected total mass enclosed within a cylinder with radius given by the average distance ($\bar{d}$ from Table \ref{tab2}) of the images from the centre of the primary lens (L1). Keeping this mass value fixed, an increment in the mass (or equivalently $\sigma_{L1}$) contribution of L1 must be compensated by a decrement in the mass (or equivalently $\sigma_{L2}$) term of L2, and viceversa. Moreover, we remark that the vicinity of the four images to L1 results in a more precise measurement of the value of $\sigma_{L1}$ (i.e., $328_{-8}^{+7}$ km s$^{-1}$). The 68\% confidence level interval (obtained by excluding from the 2000 best-fit values the 320 smallest and largest values) for $\sigma_{L2}$ gives effective velocity dispersion values between 332 and 367 km s$^{-1}$. The SDSS 3\arcsec-diameter spectrum of L2 yields an estimate of the stellar velocity dispersion of $278 \pm 6$ km s$^{-1}$ (\citealt{chr10}), that rescaled to an aperture radius of $\theta_{e}/8$ results in a central stellar velocity dispersion $\sigma_{0,L2}$ of $294 \pm 6$ km s$^{-1}$. In contrast to isolated lens galaxies, where the values of their effective and central stellar velocity dispersions are consistent within the errors (\citealt{tre06}; \citealt{gri08c}; \citealt{bol08}), the lensing measurement of the velocity dispersion of L2 is here fairly larger than the dynamical one. This can be a hint for the presence of a larger amount of dark matter in the galaxies of this system. The additional dark matter component might be associated to the haloes of the galaxies residing in an overdense environment or to a diffused dark matter halo, extended on radial scales larger than those typical of single galaxies (as in galaxy groups and clusters). We will come back to this point in the next section.

Finally, we check whether we can estimate the values of the tidal radii of the two lens galaxies, as done in recent strong lensing studies (e.g., \citealt{suy10}). We conclude that point-like images do not contain enough information to measure these physical scales. Future detailed photometric observations, providing the extended surface brightness distribution of the multiple images, would be very useful to address this last point. 

\section{The mass decomposition}

In this section we combine the results obtained above from the strong lensing modelling with those coming from a multiband photometric analysis on the lens galaxies, to infer their dark matter content. 

\begin{figure*}
  \centering
  \includegraphics[width=0.49\textwidth]{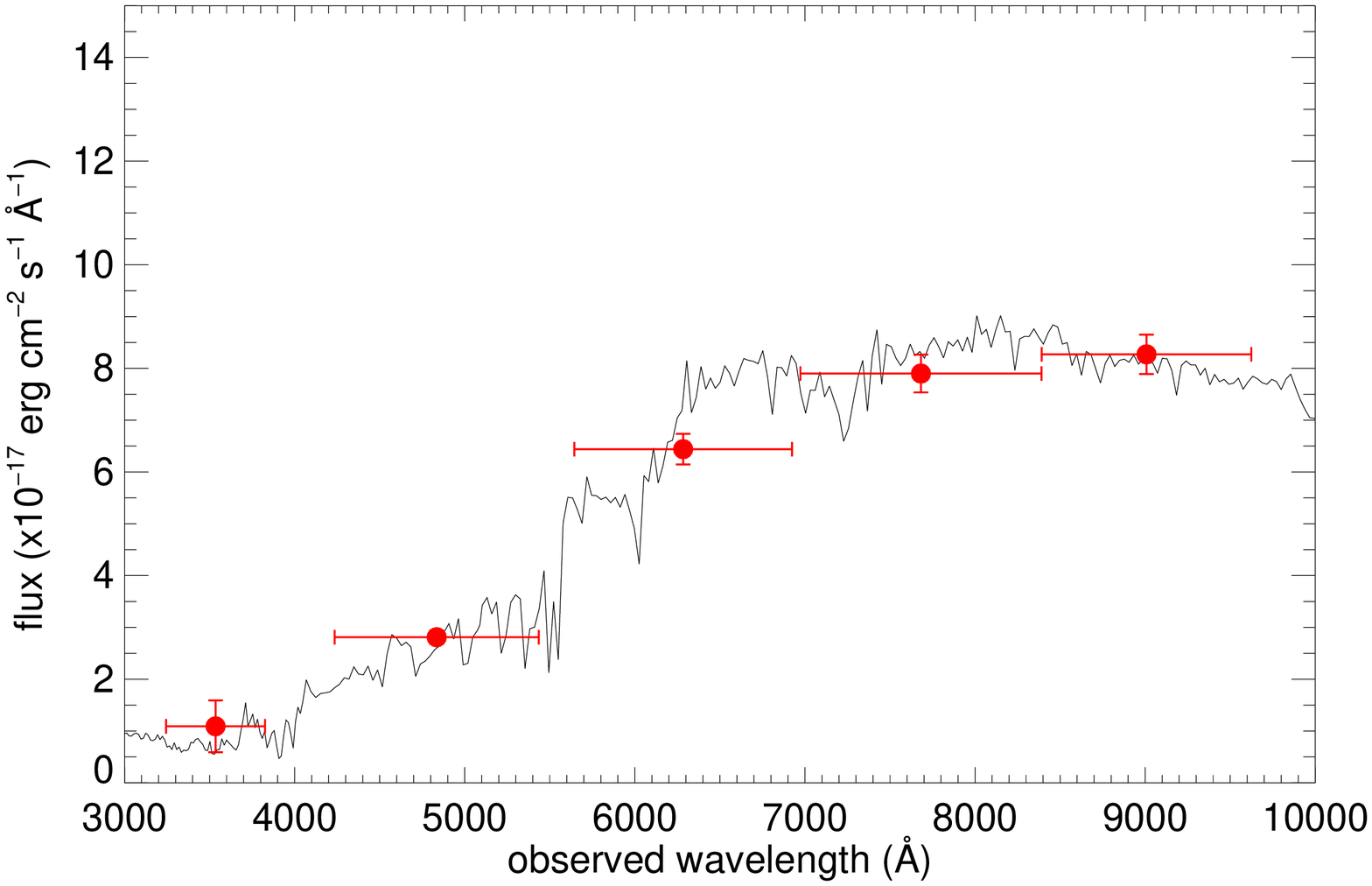}
  \includegraphics[width=0.49\textwidth]{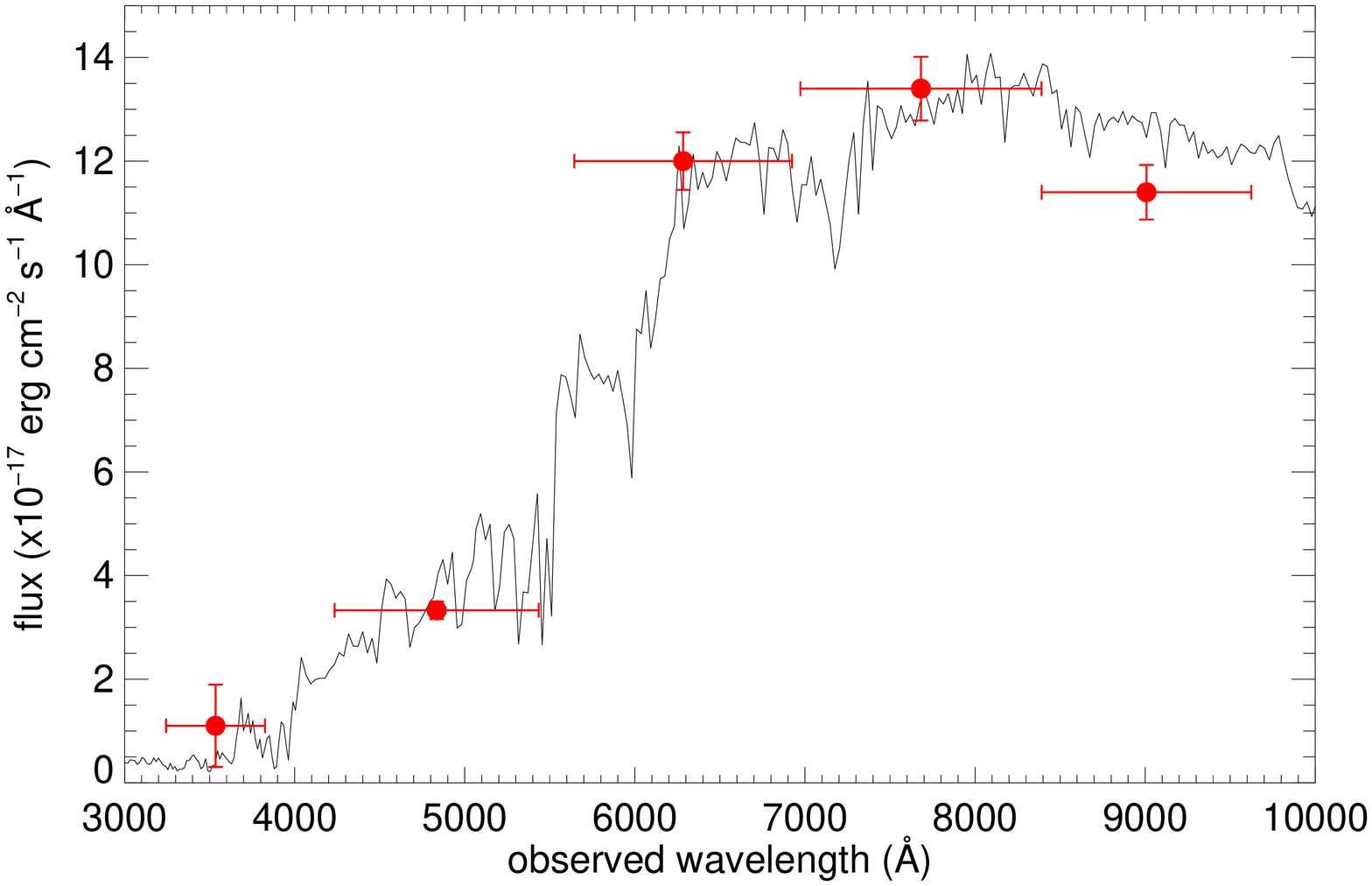}
  \caption{Spectral energy distributions and best-fit models of the lens galaxies L1 (\emph{on the left}) and L2 (\emph{on the right}). The points with the vertical error bars represent the observed total flux densities with their 1 $\sigma$ errors in the five \emph{ugriz} bands of the SDSS. The horizontal bars indicate the width of the SDSS filters. The best-fit models have been obtained using \citet{bru03} templates, a delayed exponential star formation history, and assuming a \citet{sal55} stellar initial mass function.}
  \label{fig3}
\end{figure*}

First, we estimate the composition of L1 and L2 in terms of their luminous mass. We model the five \emph{ugriz} extinction-corrected modelMag magnitudes taken from the SDSS and listed in Table \ref{tab1} with composite stellar population models, by using the photometric redshift code HyperZ (\citealt{bol00}). For the fits, we fix the redshifts of L1 and L2 to the spectroscopic value measured for L2 (see Table \ref{tab1}). We adopt \citet{bru03} templates with a delayed exponential star formation history (for more details on the method and on its reliability, see \citealt{gri08a,gri09,gri10a}). Following the results of several recent studies on massive early-type galaxies (e.g., \citealt{gri08a,gri09}; \citealt{gri10b}; \citealt{tre10}; \citealt{aug10}), we adopt a \citet{sal55} stellar initial mass function. We find that the total luminous mass values for L1 and L2 are $(3.09\pm0.30)\times 10^{11}$ and $(5.87\pm0.58)\times 10^{11}$ $M_{\odot}$, respectively. These correspond to stellar mass-to-light ratio values in the observed $i$-band of $2.5\pm 0.3$ and $2.8 \pm 0.3$ $M_{\odot}L_{\odot,i}^{-1}$. The galaxy spectral energy distributions with the best-fit models are plotted in Fig. \ref{fig3}.

\begin{figure*}
  \centering
  \includegraphics[width=0.49\textwidth]{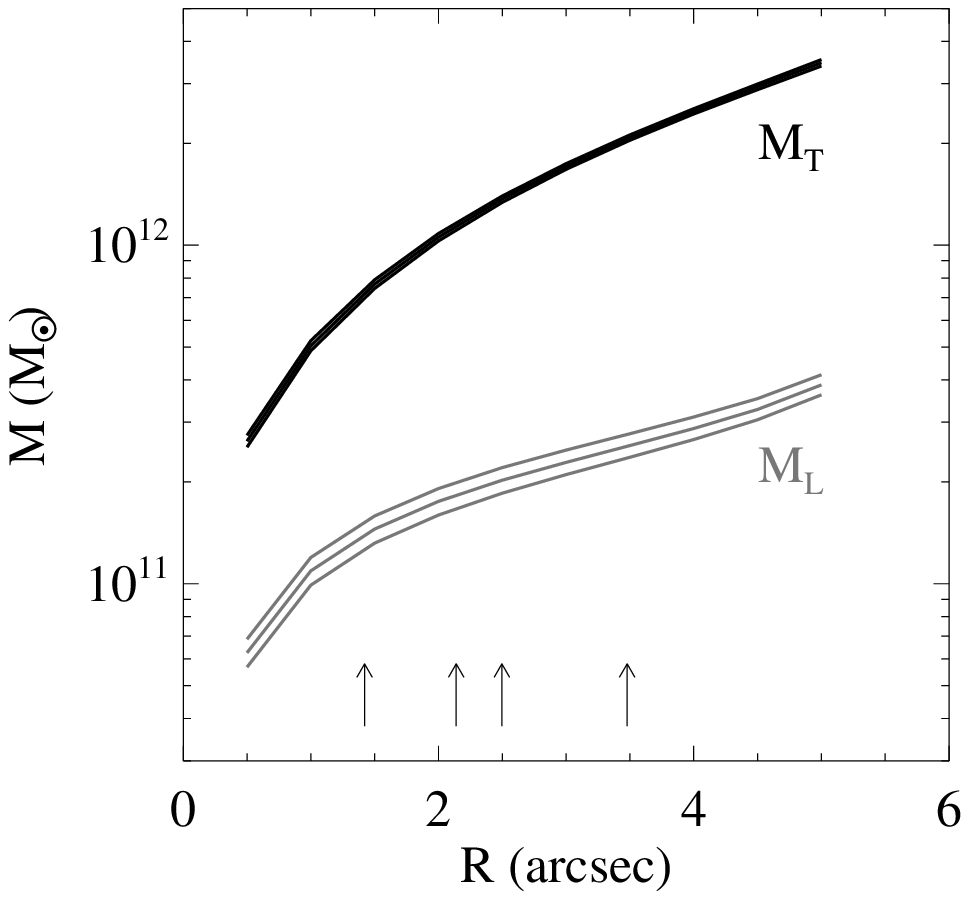}
  \includegraphics[width=0.49\textwidth]{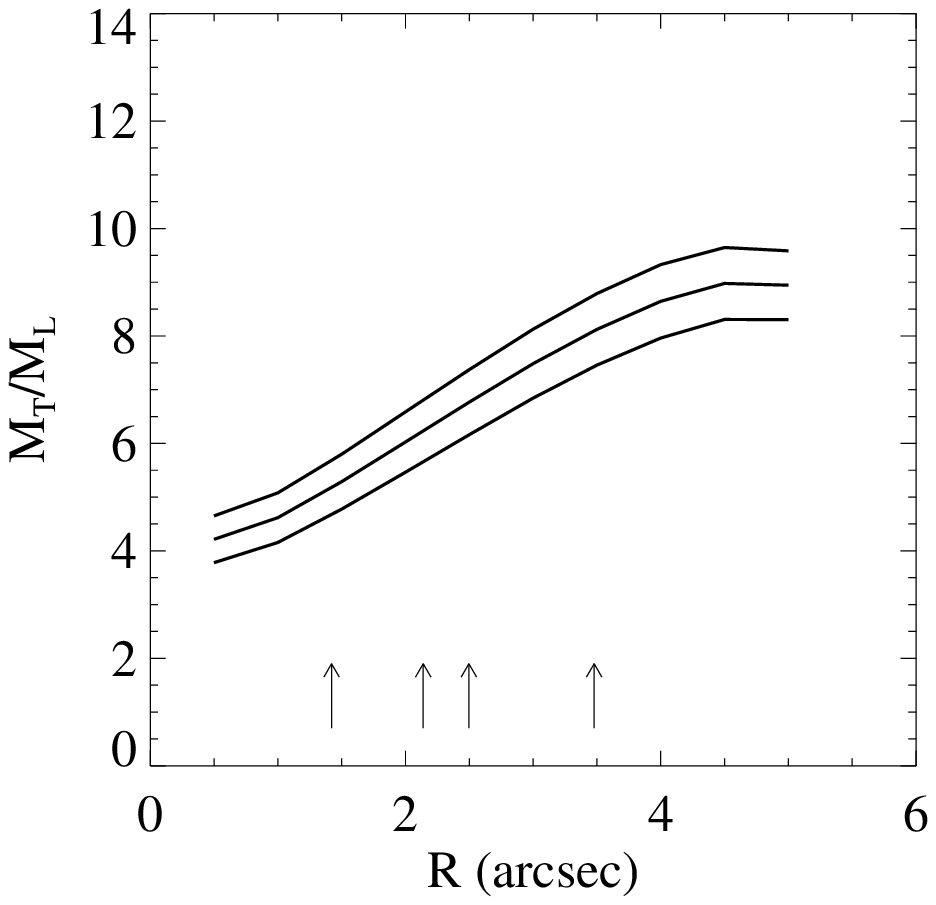}
  \caption{Projected cumulative luminous and total mass distributions. The arrows locate the projected angular distances of the four observed multiple images from the lens galaxy L1. \emph{On the left:} The luminous and total mass components (with their 1 $\sigma$ confidence regions), derived, respectively, from strong gravitational lensing and stellar population analyses. \emph{On the right:} The ratio of the total over luminous mass with the 1 $\sigma$ errors, related to the amount of dark matter present in the deflector.}
  \label{fig4}
\end{figure*}

Then, we simulate 2000 luminous mass distributions for the two lens galaxies by choosing de Vaucouleurs profiles and taking into account the errors on the total luminous masses. We measure the projected luminous mass values inside cylinders centred on L1 and with radii going from 0.5 to 5\arcsec\, by adding over these regions the contributions from L1 and L2. We obtain that inside the average distance of the four multiple images from L1 the projected luminous mass is $M_{L}(R<2.38\arcsec)=1.96_{-0.17}^{+0.18}\times 10^{11}M_{\odot}$. The projected cumulative luminous mass with the 1 $\sigma$ confidence level errors is shown in Fig. \ref{fig4}.

Next, by using the results coming from the bootstrapping analysis described in the previous section, we determine the projected total mass values inside the same aperture radii defined for the projected luminous mass estimates. We employ the 2000 best-fit values of the $b_{L1}$ and $b_{L2}$ parameters to measure the projected total mass distributions of the two lens galaxies and sum their contributions. We find that inside an aperture of 2.38\arcsec\, the projected total mass is $M_{T}(R<2.38\arcsec)=(1.30\pm 0.03)\times 10^{12}M_{\odot}$. The projected cumulative total mass with the 1 $\sigma$ confidence level errors is shown in Fig. \ref{fig4}.

In Fig. \ref{fig4}, we also plot the projected total to luminous mass ratio. This quantity is significantly larger than one at all radii, implying the presence of a remarkable dark matter component in addition to the luminous one. In detail, at the average distance of the multiple images from L1 the fraction of projected dark over total mass is $f_{D}(R<2.38\arcsec)=0.8\pm0.1$. An angular distance of 2.38\arcsec\, from the lens centre corresponds to approximately 1.3 times the value of the galaxy effective angle (see Table \ref{tab1}). At this rescaled distance ($R = 1.3 \times R_{e}$), a high value of $f_{D}$ such as that found in this lens galaxy is extremely uncommon in isolated lens (and non-lens) massive early-type galaxies (see \citealt{gav07}; \citealt{gri09}; \citealt{gri10a}). This fact, combined with the observational evidence of a statistically significant amount of dark matter already at distances smaller than 1\arcsec$\ $ from the centre of L1, supports the picture that the deflector of this system is a galaxy group. Deeper multiband images, with a better angular resolution, would offer the opportunity of testing whether, as in galaxy clusters, the dark matter component is distributed on a spatial scale larger than that typical of single galaxies. 

A different way of establishing the presence of a large amount of dark matter in the central regions of these galaxies consists in comparing the values of stellar and total mass-to-light ratios within the galaxy effective radii. We measure projected total (stellar and dark) mass-to-light ratios of $12.6 \pm 1.4$ and $13.1 \pm 1.7$ $M_{\odot}L_{\odot,i}^{-1}$, respectively, for L1 and L2. These values are significantly larger than the stellar mass-to-light ratios reported above (and typical of early-type galaxies). From this result, we can further exclude at a very high statistical level that the lens galaxies are composed of only luminous matter.

\begin{figure}
  \centering
  \includegraphics[width=0.49\textwidth]{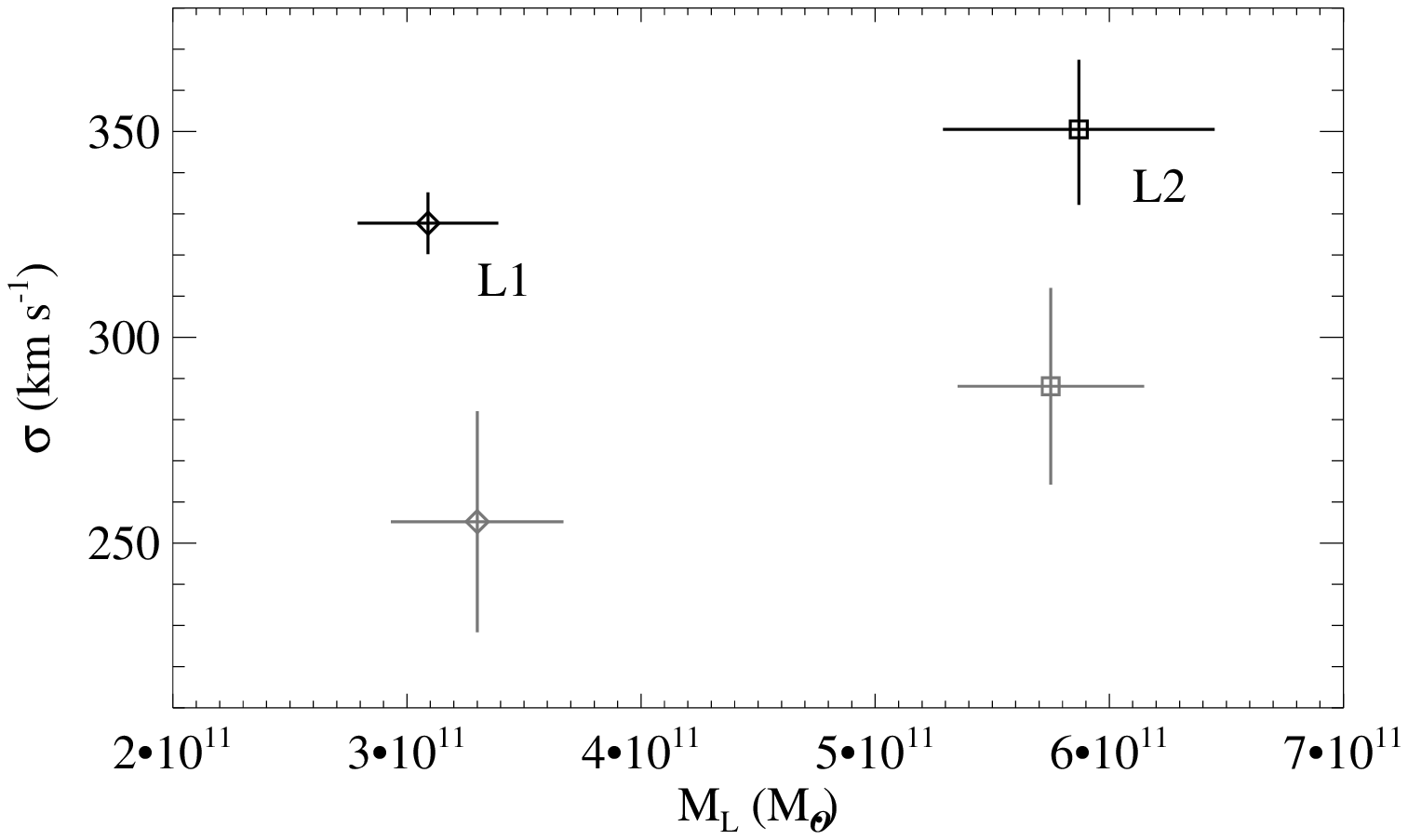}
  \includegraphics[width=0.49\textwidth]{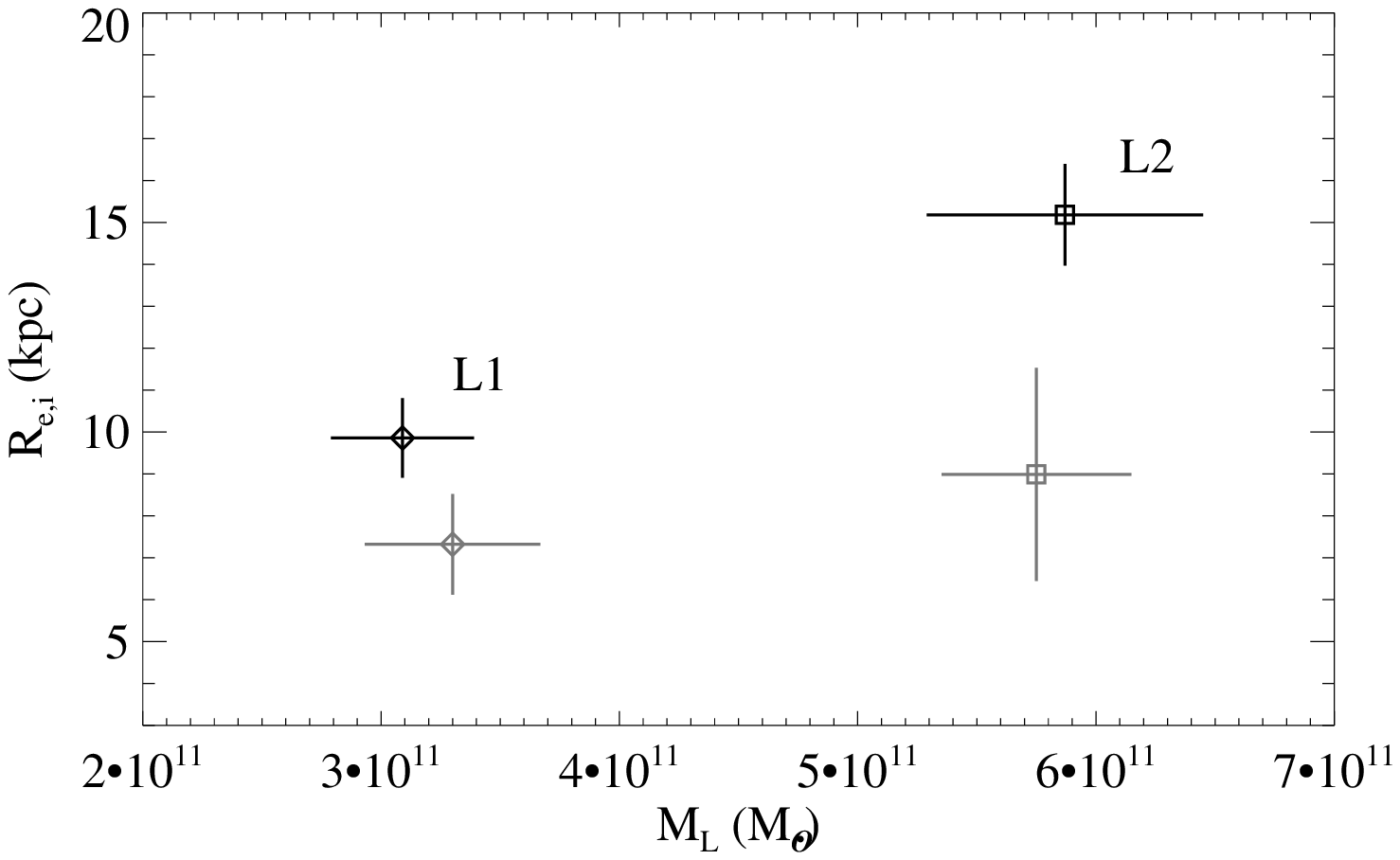}
  \includegraphics[width=0.49\textwidth]{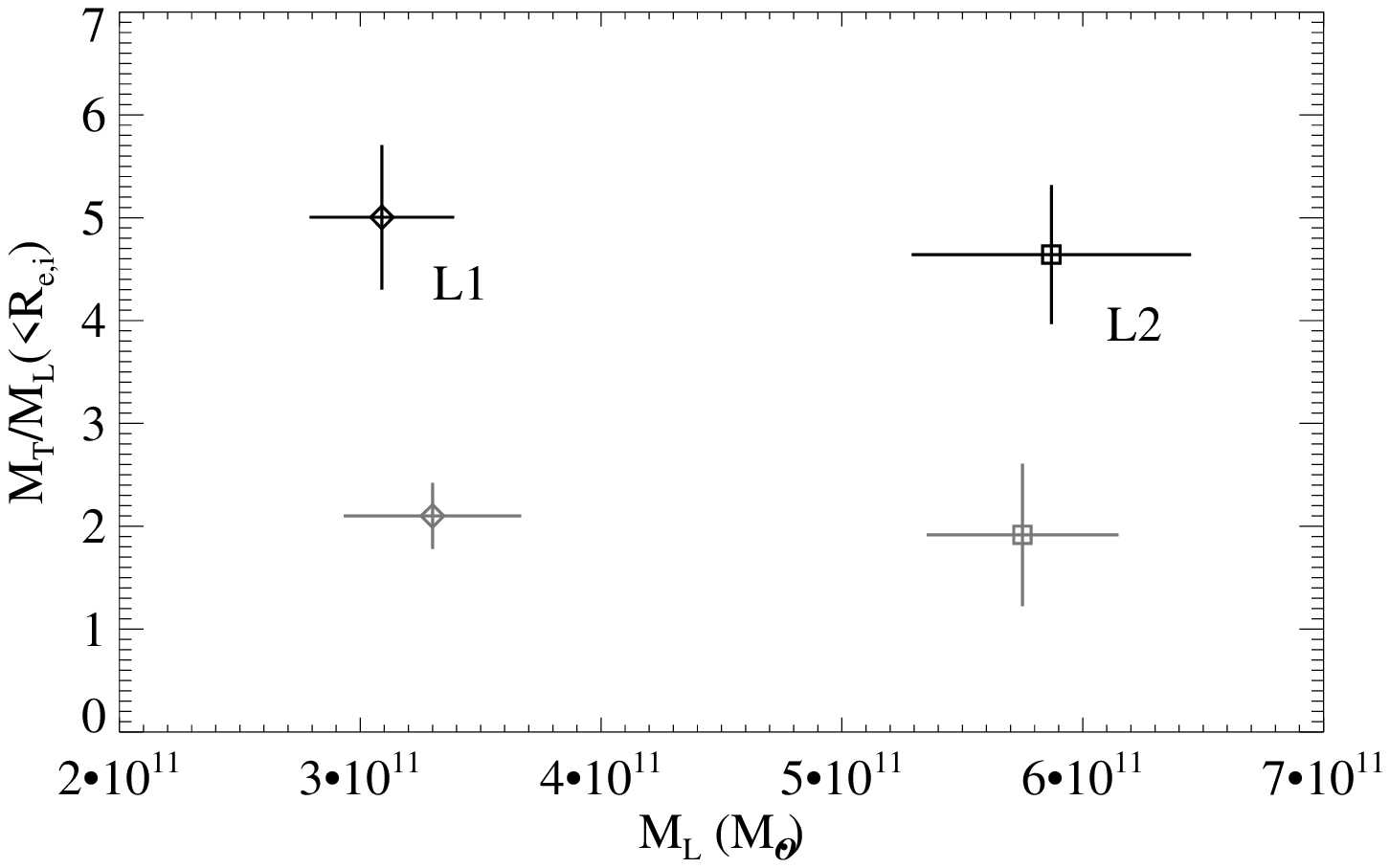}
  \caption{CASSOWARY 5 versus field lens galaxies. The physical quantities for L1 and L2 are indicated in black and for the two samples of 10 lens galaxies from the SLACS survey in grey. The values of the effective velocity dispersions of the isothermal lensing models $\sigma$ (\emph{top panel}), the effective radii in the L1-L2 \emph{i}-band local frame $R_{e,i}$ (\emph{middle panel}), and the projected total over luminous mass ratios within the effective radius $M_{T}/M_{L}(<R_{e,i})$ (\emph{bottom panel}) are plotted as a function of the luminous mass values. The error bars for the SLACS samples are the simple standard deviation values of the points and not the errors on the mean values.}
  \label{fig6}
\end{figure}

To investigate possible galaxy structural differences related to the environment in which lenses reside, we consider two samples, each composed of 10 galaxies, from the SLACS survey with the 10 closest luminous mass values (taken from \citealt{gri09}) to those of L1 and L2. \citet{tre09} have shown that although the SLACS galaxies (being mainly massive early-type galaxies) prefer overdense environments, less than 20\% of them belong to a group or a cluster. In the same study, it has also been measured that the contribution of the environment to the potential of the SLACS lenses is no more than a few per cent. In Fig. \ref{fig6}, we contrast the average physical scales of the typically field lens galaxies in the two SLACS samples with those of the lens galaxies L1 and L2. We show that at fixed luminous mass the values of the effective velocity dispersions of the best-fit isothermal models in these lens galaxies are significantly larger than in field lens galaxies. We also observe that the values of the effective radii measured in the \emph{i}-band reference frame at the redshift of L1 and L2 are for these lens galaxies larger than for field lens galaxies with similar luminous masses. Moreover, we find that the ratio between the total (estimated by assuming an isothermal model) and luminous masses projected within a cylinder with radius equal to the effective radius is considerably larger in these than in field lens galaxies, at comparable luminous masses.

The third panel of Fig. \ref{fig6} proves quantitatively that L1 and L2 contain, within their effective radii, an approximately factor four larger amount of dark matter than field lens galaxies with similar luminous masses. In passing, we mention that a much larger sample of non-lens massive early-type galaxies selected from the SDSS show projected dark over total mass fractions consistent with those of the SLACS lenses plotted here (see \citealt{gri10a}). The first and second panel of the same figure suggest that not only the total but also the luminous properties of the galaxies are significantly affected by the different environmental characteristics and concentration of dark matter. A rate of dry minor mergers larger in a group than in the field environment could be at the origin of the different physical properties of the galaxies (for some additional results on the importance of these evolutionary mechanisms in massive early-type galaxies, see \citealt{gri10a}).

\section{Conclusions}

We have investigated the gravitational lensing system CSWA 5 composed of two massive early-type galaxies at redshift 0.388 acting as main lenses on a source at redshift 1.069. Four lensed images of this source are observed at an average angular distance from the primary lens galaxy of 2.38\arcsec. By modelling the total mass distributions of the two galaxies with singular isothermal sphere profiles, we have shown that the geometry of the multiple image system can be accurately reproduced. We have then combined the total mass estimates provided by the lens modelling with the luminous mass measurements obtained by fitting with composite stellar population models the SDSS multicolor photometry of the lens galaxies. This has allowed us to explore the dark matter content of these galaxies. The main results of this study can be summarized as follows:
\begin{itemize}

\item[$-$] The projected total mass distribution of the two lens galaxies is described better by isothermal than de Vaucouleurs profiles, showing that the galaxy projected total mass is less concentrated than light. 

\item[$-$] The effective velocity dispersion values of the isothermal models of L1 and L2 are anticorrelated, can be measured with a good precision, and are $328^{+7}_{-8}$ and $350^{+17}_{-18}$ km s$^{-1}$, respectively. The latter is larger than the galaxy central stellar velocity dispersion value ($294 \pm 6$ km s$^{-1}$).

\item[$-$] According to the best-fit lens model, each observed multiple image is magnified by a factor larger than 2 and the total magnification of the lensed source is approximately 17.

\item[$-$] Assuming a Salpeter stellar initial mass function, the modelling of the galaxy spectral energy distributions provides luminous mass values of $(3.09\pm0.30)\times 10^{11}$ and $(5.87\pm0.58)\times 10^{11}$ $M_{\odot}$ for L1 and L2, respectively.

\item[$-$] The total and luminous masses projected within a cylinder centred on L1 and with radius of 12.6 kpc, given by the average distance of the multiple images from L1, are $(1.30\pm 0.03)\times 10^{12}$ and $1.96_{-0.17}^{+0.18}\times 10^{11}$ $M_{\odot}$, respectively. This implies a projected dark over total mass ratio of approximately $0.8 \pm 0.1$ at an angular distance of only 1.3 times the value of the effective radius of L1.

\item[$-$] In the observed $i$-band, the values of the stellar mass-to-light ratio of L1 and L2 are, respectively, $2.5 \pm 0.3$ and $2.8 \pm 0.3$ $M_{\odot}L_{\odot,i}^{-1}$. Within the effective radii of L1 and L2, the values of the projected total mass-to-light ratio are $12.6 \pm 1.4$ and $13.1 \pm 1.7$ $M_{\odot}L_{\odot,i}^{-1}$, respectively. This excludes, at a very high confidence level, a composition of the two galaxies in terms of luminous mass only.

\item[$-$] The galaxies L1 and L2 show an amount of dark matter in their internal regions almost four times larger than in field lens galaxies with similar luminous masses. This fact is consistent with both galaxies residing in a galaxy group. The effective radii of these two galaxies are also larger than in comparable isolated lens galaxies. These results provide some observational evidence on the different galaxy evolution processes taking place in overdense and more dark matter-concentrated environments.

\end{itemize}

A deeper image with a better angular resolution could provide interesting information on the model-predicted presence of a fifth demagnified image, on the values of the tidal radii of the lens galaxies, and on the radial scale of the dark matter distribution. Our findings offer interesting information about the physical processes that determine the formation of structures in the Universe. Extending studies of this kind to more strong lensing systems living in disparate environments and comparing the observational results with those coming from (baryons+dark matter) cosmological simulations would greatly help us understanding the intricate mechanisms that are responsible for the mass assembly and structural evolution of galaxies.    

\section*{Acknowledgments}

We thank Matthew Auger, Raphael Gobat, and Tommaso Treu for useful discussions. This research was supported by the DFG cluster of excellence ``Origin and Structure of the Universe''. 

This work has made extensive use of the SDSS database. Funding for the SDSS and SDSS-II has been provided by the Alfred P. Sloan Foundation, the Participating Institutions, the National Science Foundation, the U.S. Department of Energy, the National Aeronautics and Space Administration, the Japanese Monbukagakusho, the Max Planck Society, and the Higher Education Funding Council for England. The SDSS Web Site is http://www.sdss.org/. The SDSS is managed by the Astrophysical Research Consortium for the Participating Institutions. The Participating Institutions are the American Museum of Natural History, Astrophysical Institute Potsdam, University of Basel, University of Cambridge, Case Western Reserve University, University of Chicago, Drexel University, Fermilab, the Institute for Advanced Study, the Japan Participation Group, Johns Hopkins University, the Joint Institute for Nuclear Astrophysics, the Kavli Institute for Particle Astrophysics and Cosmology, the Korean Scientist Group, the Chinese Academy of Sciences (LAMOST), Los Alamos National Laboratory, the Max-Planck-Institute for Astronomy (MPIA), the Max-Planck-Institute for Astrophysics (MPA), New Mexico State University, Ohio State University, University of Pittsburgh, University of Portsmouth, Princeton University, the United States Naval Observatory, and the University of Washington.


\label{lastpage}

\end{document}